\documentclass[12pt]{article}
\usepackage[hmargin=2.2cm,vmargin=2.5cm]{geometry}
 \usepackage{graphicx}
 \usepackage{amsmath}
 \usepackage{amssymb}
  \usepackage{enumerate}
\usepackage{cite}
\usepackage{mathabx}
\usepackage{url}
\newcommand{\bey}{\begin{eqnarray}}
\newcommand{\eey}{\end{eqnarray}}
\usepackage{etoolbox}

\usepackage{lipsum}

\let\OLDthebibliography\thebibliography
\renewcommand\thebibliography[1]{
  \OLDthebibliography{#1}
  \setlength{\parskip}{0pt}
  \setlength{\itemsep}{2pt plus 0.3ex}
}

\newcommand{\R}{\mathbb{R}}

\newcommand{\po}{\prec}

\begin{document}
\title{Quantum probabilities for the  causal ordering of events}
\author {Charis Anastopoulos\footnote{anastop@upatras.gr}  \\
 {\small Department of Physics, University of Patras, 26500 Greece} \\
  and   \\
  Maria-Electra Plakitsi \\
 {\small Department of Electrical and Computing Engineering, University of Patras, 26500 Greece }}
\maketitle
\begin{abstract}
We develop a new formalism for constructing probabilities associated to  the causal ordering of events in quantum theory, where by an event we mean the emergence of a measurement record on a detector. We start with constructing probabilities for the causal ordering events in classical physics, where events are defined in terms of worldline coincidences. Then, we show how these notions generalize to quantum systems, where there exists no fundamental notion of trajectory. The probabilities constructed here are experimentally accessible, at least in principle. Our analysis here clarifies  that the existence of quantum orderings of events does not require quantum gravity effects: it is a consequence of the quantum dynamics of matter, and it appears in presence of a fixed background spacetime.
\end{abstract}

\section{Introduction}
A bet on any type of  race (with humans, horses, chariots, or cars) is equivalent to the assignment of probabilities to a causal ordering of events. The relevant events  are the crossings of the finish line by the runners, and the causal ordering of such events is the results of the race. In this sense, assigning probabilities to causal orderings is both one of the oldest applications of probabilistic thinking, dating at least to the ancient Olympics, and one of the most common  uses of probability theory today.
In this paper, we describe causal ordering of events (COoE) for quantum systems, where by an event we mean the emergence of a macroscopic measurement record that is localized in space and in time \cite{QTP1, QTP6}. We construct the probabilities for such causal orderings, and we suggest physical set-ups where such probabilities can be measured, at least in principle.

This work is partially motivated by the recent studies of indefinite causal ordering of events in quantum computing \cite{OCB12, CDPV13}. In this context, the word ``event" is used to denote an operation on a quantum system, for example, a step in an algorithm. An indefinite sequence of operations can arguably lead to significant advances in quantum computation and other technologies \cite{ca1, ca2, ca3, ca4}. The most common set-up to witness such phenomena involves the quantum switch, that is, a quantum operation in which two or more quantum channels act on a quantum system with the order of application determined by the state of another quantum system. Systems that manifest indefinite causal order in this sense have been realized in the laboratory \cite{exp1, exp2, GR20}. 

This quantum informational notion of causal ordering differs and may even conflict with relativistic causality---for a detailed analysis of this issue, see \cite{ViCo, ViRe}. The meaning of the term ``event" is crucial in this context. In this paper, we employ a notion of event that is similar to the crossing of the finishing line by a racer. This notion closely reflects the notion of an event in relativity, where  physical events are invariantly defined in terms of world-line coincidences\footnote{Einstein emphasized this perspective in his first review of General Relativity \cite{Einstein16}, see also \cite{Bergmann} and \cite{Mermin}.}. For example, a particle-detection event is defined as the intersection of the particle's and the detector's worldlines. With this definition, we use the lightcone structure of spacetime, in order to define causal relations between events.  Since events are defined in terms of trajectories, the dynamical behavior of trajectories directly influences the properties of causal ordering. If the trajectories are stochastic, then the ordering of events is also a stochastic variable. 

Quantum theory does not admit trajectories as physical observables, so in quantum systems, we have to define events and their causal ordering differently. We define an event as the occurrence of a measurement outcome, that is, the emergence of a measurement record on a detector. This is the most conservative definition of a quantum event. It is also the most natural one in the Copenhagen interpretation \cite{Peres}, and the standard use of the term in particle physics.
 If the quantum events  can be embedded in spacetime, i.e., if we can associate a spacetime point or region to the emergence of a measurement record, then we can define quantum probabilities for the COoE that are natural analogues of the  classical ones.

Such a definition requires the treatment of the time associated to an event as a quantum observable. It is an old result by Pauli \cite{Pauli} that time cannot be treated as a self-adjoint operator. The only way to have time as an observable is to represent it by a Positive-Operator-Valued measure (POVM). To this end, we use the Quantum Temporal Probabilities (QTP) approach that has been developed for constructing probabilities for temporal observables \cite{QTP1, QTP2, QTP3,  QTP5, QTP6}.

The key idea  in QTP is to distinguish between the time parameter of Schr\"odinger's equation and the time variable associated to  particle detection \cite{Sav99, Sav10}. The latter is then treated as  a macroscopic quasi-classical variable associated to the detector degrees of freedom. A quasi-classical variable is a coarse-grained quantum variable that  satisfies appropriate decoherence conditions, so that its time evolution can be well approximated by  classical equations  \cite{GeHa2, Omn1, Ana23}.  Hence, the detector admits a dual description: in microscopic scales it is described   by quantum theory, but its macroscopic records are expressed  in terms of classical spacetime coordinates. The key point here is that the spacetime coordinates of a quantum event are random variables, and they can be used to  define quantum probabilities for the causal order of events. 

It is important to emphasize  that there is  no relation between the COoE considered here and a quantum causal structure of spacetime, as commonly  postulated in quantum gravity research. The quantum behavior of the COoE, considered here, is due to the quantum nature of matter, and it coexists peacefully with a fixed background spacetime. In fact, the background spacetime structure is essential for defining quantum probabilities for the COoE.

We also show that probabilities for quantum COoEs can also be defined even if there is no macroscopic record about the time at which the events occur. To this end, we construct a simple detection model, in which different orderings of events correspond to different measurement records. Hence, the probabilities of such records coincide with probabilities for different causal orders. 

The structure of this paper is the following. In Sec. 2, we provide a general definition of the notion of an event, and explain how we can construct probabilities for the causal order of events. In Sec. 3, we apply these definitions to classical physics, including Hamiltonian mechanics and stochastic processes. In Sec. 4, we define probabilities for the causal ordering of quantum events, using temporal observables. In Sec. 5, we present a simple model for the quantum order of events in absence of records about temporal observables. In Sec. 6, we summarize   our results.

\section{Main concepts}
In this section, we  present a general characterization of events, and we identify the mathematical  properties that are satisfied by a causal order of events.

 By ``event" we mean a uniquely identifiable   occurrence with definite characteristics. In classical  mechanics, events are typically defined as the intersection of two world-lines. For example, one worldline may correspond to a particle and the other to an observer with a particle detector; their coincidence is a particle-detection event. We can improve on this description, by defining an event as  the first intersection of a worldline with a specific time-like surface. We can consider, for example, the worldline of a runner crossing the world tube of the finish line in a marathon. 

However, in quantum theory, definite characteristics are attributed only to measurement outcomes; trajectories are not  observables. For this reason, we will define events in terms of measurement records. For example, a detection event is the ``click" of a particle detector. Hence, we identify event with changes in a macroscopic apparatus that denote the occurrence of a measurement.

Let $E$ be a set of events in the physical system under study. Events are discrete occurrences, so $E$ is a discrete set. We will denote events by Greek letters, $\alpha, \beta$, and so on.

The events in the set $E$ may be ordered causally. We say that $\alpha \po \beta$, if event $\alpha$ occurs prior to an event $\beta$. A {\em causal order} on $E$ is the consistent assignment of the order relation $\po$ to pairs of elements of $E$. A causal order satisfies the properties of a partial-order relation, namely, 
\begin{enumerate}
\item    Irreflexivity: It is never true that $\alpha \po \alpha$.
  \item   Asymmetry: If $\alpha \po \beta$, then $\beta \po \alpha$ is false.
 \item   Transitivity: If $\alpha \po \beta$ and $\beta \po \gamma$, then $\alpha \po \gamma$. 
 \end{enumerate}

 In a partial order, it is not necessary that all pairs of elements are related with the order relation. Physically, we can distinguish two cases. Some elements may be {\em simultaneous}, in which case we write $\alpha \sim \beta$. Or they may be {\em uncomparable}, in which case we write $\alpha | \beta$. We therefore define a causal order as a partial order that also  include the distinction between simultaneous and incomparable pairs of elements. 
 
 We will denote the set of all possible causal orders on $E$ by  $CO(E)$. We will denote elements of $CO(E)$ by capital Greek letters. For example, in a set $E$ that consists of two distinct elements $\alpha$ and $\beta$, there are four possible causal orders
 \begin{itemize}
  \item $M_1 = \{ \alpha \po \beta \} $
  \item  $M_2 = \{ \beta \po \alpha, \} $
  \item $M_3 = \{\alpha | \beta\} $
  \item $M_4 = \{ \alpha \sim \beta\}$
\end{itemize} 
  
  We say that a causal order defines a {\em time order} on $E$,  if there exist no pair  $\alpha, \beta \in E$ such that $\alpha | \beta$. We will denote the set of all time orders on $E$ by $TO(E)$. Clearly $TO(E) \subset CO(E)$. 
 
Ever since Newton, we define the causal ordering of physical events in terms of  the spacetime causal structure. That is, we consider a four-dimensional manifold $M$ with points $(x^0, x^1, x^2, x^3)$ that is equipped with a partial ordering relation $<$ that defines the causal structure of spacetime.

\begin{itemize}
\item In non-relativistic physics, $x < y$ if $x^0 < y^0$, and  $ x \sim y$ if $x^0 = y^0$. There are no incomparable elements.  
\item In relativistic physics, $ x < y$, if $y$ lies in the future lightcone of $x$, and $x|y$ if $x$ is spacelike separated from $y$. Spacelike separated events are incomparable; there are no simultaneous events. 
\end{itemize}

Since all physical events occur in spacetime, we consider embeddings $X$ of the set of events $E$ into spacetime, that is, onto maps $X: E \rightarrow M$. Then, the  pullback of the spacetime causal structure with respect to $X$ defines a causal order on $E$, that is,
\bey
\alpha \po \beta, \ \ \text{if} \ \ X(\alpha) < X(\beta)
\eey
Hence, the physical COoE reflects the causal structure on spacetime\footnote{Here, we associated events with spacetime points. A more general analysis should associate events with spacetime {\em regions}, but it will not be needed in  this paper. }. 

It is imperative to distinguish COoEs from the causal structure of spacetime. As long as we ignore gravitational interactions, the latter is fixed an unchanging. It is defined by the lightcone structure of Minkowski spacetime, or of any other background spacetime.
 However, COoEs are not fixed: they  can be  stochastic or quantum variables. This is because they depend on the embeddings $X$, which are themselves stochastic or quantum variables. The quantum behavior of the COoEs does not require a quantum behavior of spacetime, 
 as postulated in quantum gravity theories. As a matter of fact, in quantum gravity proper, we expect to have no external spacetime causal structure, hence, the definitions of the COoEs given here do not work. The difficulties that arise from this fact are known as the {\em problem of time} in quantum gravity \cite{Isham92, Kuchar92, Anderson17}. 
 
 \section{COoE in classical physics}
 In this section, we construct probabilities for the COoEs for classical systems, namely, for Hamiltonian systems and for systems described by stochastic processes.
 
 \subsection{Classical mechanics }
Let $\Gamma$ be the state space of a classical system; we will denote the elements of $\Gamma$ by $\xi$. By Hamilton's equation, a system found at $\xi$ at time $t = 0$, will evolve to a point $\sigma_t(\xi)$ at time $t$. The map $\sigma_t$ is a diffeomorphism.

An event in classical mechanics corresponds to the first intersection of a state space trajectory with a surface  on $\Gamma$. Surfaces of codimension $s$ are locally determined by the vanishing of $s$ functions on $\Gamma$, hence, we can represent an event $\alpha$ by a set of $s$ functions $F^i_{\alpha}$, where $i = 1, 2, \ldots s$. A set $E$ of $n$  events consists $n$ such families. For simplicity, we will consider only surfaces of codimension one in this paper, so that one event corresponds to a single function on $\Gamma$.
 
The causal ordering in the set of events is defined through the parameter of time evolution $t$, which is assumed to coincide with the Newtonian absolute time. For any event $\alpha$, we define the null set $N_{\alpha}$ of $\alpha$, as the set of all $\xi \in \Gamma$, such that the equation $F_{\alpha}[\sigma_t(\xi)] = 0$ has no solution for all $t \geq 0$. Then, we define the time $T_{\alpha}$ of the event $\alpha$, as a function $T_{\alpha}: \Gamma - N_{\alpha} \rightarrow \R^+ $, such that $T_{\alpha}(\xi)$ is the smallest positive value of $t$ that solves the equation $F_{\alpha}[\sigma_t(\xi)] = 0$. This means that $T_{\alpha}(\xi)$ is the time it takes a trajectory that starts at $\xi$ to cross the surface $F_{\alpha} = 0$ for the first time.  

Suppose that the initial state of the system corresponds to a probability distribution $\rho(\xi)$. Then, we can construct joint probability distributions for the times of events
\bey
p(t_1, t_2, \ldots, t_n) = \int d\xi \rho(\xi) \delta[T_1(\xi) - t_1] \delta[T_2(\xi) - t_2] \ldots \delta[T_n(\xi) - t_n] . 
\eey
These probability densities are not normalized to unity. For proper normalization, we have to include the probability densities for no events, which corresponds to the null sets $N_{\alpha}$. For example, for $n = 2$, we have the probability densities $p(t_1, t_2)$ as above, together with the probability densities
\bey
p(N_1, t_2) = \int d \xi \chi_{N_1}(\xi) \delta[T_2(\xi) - t_2] \\
p(t_1, N_2) = \int d \xi \chi_{N_2}(\xi) \delta[T_1(\xi) - t_1]\\
p(N_1, N_2) = \int d\xi \chi_{N_1}(\xi)\chi_{N_2}(\xi).
\eey
 Here, $\chi_C$ is the characteristic function of a set $C$.
 
We can define the following four causal orders for the two events, 
\begin{itemize}
\item $M_1 = \{1 \po 2\}$ corresponds to $t_1 < t_2$, or $N_2$ together with finite $t_1$.
\item  $M_2 = \{2 \po 1\}$ corresponds to $t_2 < t_1$, or $N_1$ together with finite $t_2$.
\item $M_3 = \{1 | 2\}$ corresponds to $N_1$ and $N_2$.
\item $M_4 = \{1 \sim 2\}$ corresponds to $t_1 = t_2$. 
\end{itemize}
Then, we obtain the associated probabilities  
\bey
p(M_1) &=& \int_0^{\infty} dt_1 \int_0^{t_1} dt_2 p(t_1, t_2) + \int_0^{\infty} dt_1 p(t_1, N_2) \nonumber \\
p(M_2) &=& \int_0^{\infty} dt_2 \int_0^{t_2} dt_1 p(t_1, t_2) + \int_0^{\infty} dt_2 p(N_1, t_2) \nonumber \\
p(M_3) &=& p(N_1, N_2) \nonumber \\
p(M_4) &=& \int_0^{\infty} dt p(t, t).
\eey
This procedure is straightforwardly generalized to $n$ events. 

As an illustration, consider a system of two free  particles of mass $m$ in one dimension, with state space $\Gamma = \{ x_1, x_2, p_1, p_2\}$. We restrict to $ x_1 \leq 0$ and $x_2 \leq 0$, and we consider the pair of events that correspond to either of the two particles crossing the line $x = 0$. Hence, the two functions that define events are $F_1 = x_1$ and $F_2 = x_2$. The equations of motion are $x_1(t) = x_1 + p_1t/m$ and $x_2(t) = x_2 + p_2 t/m$. We straightforwardly find that $N_1 = \{(x_1, x_2, p_1, p_2)|p_1 \leq 0\}$, and $N_2 =  \{(x_1, x_2, p_1, p_2)|p_2 \leq 0\}$, that is, the particles never cross the line $x = 0$ if they have non positive momentum. Similarly, we compute the time functions $T_1 = - mx_1/p_1$ and $T_2 = - mx_2/p_2$. 

It is simple to identify the subsets of $\Gamma$ that correspond to the different causal orders,
\bey
M_1 &=& \{(x_1, x_2, p_1, p_2)|p_1 > 0, p_2 >0,  x_1p_2 > x_2 p_1\} \cup \{(x_1, x_2, p_1, p_2)|p_1 > 0, p_2 \leq 0 \},\nonumber \\
M_2 &=& \{(x_1, x_2, p_1, p_2)|p_1 > 0, p_2 >0,  x_2p_1 > x_1 p_2\} \cup \{(x_1, x_2, p_1, p_2)|p_1 \leq 0, p_1 > 0 \}, \nonumber \\
M_3 &=& \{ (x_1, x_2, p_1, p_2)| p_1 \leq 0, p_2 \leq 0\}, \nonumber \\
M_4 &=& \{(x_1, x_2, p_1, p_2)|p_1 > 0, p_2 >0,  x_2p_1 = x_1 p_2\}. 
\eey
The associated probabilities are simply $p(M_i) = \int d\xi \chi_{M_i}(\xi) \rho(\xi)$.
Note that $M_4$ is a set of measure zero, so, in general, the associated probability vanishes. 

Suppose, for example, that both particles start from $x_0 < 0$, and that they have the same momentum distribution $f(p)$, so that 
\bey
\rho(x_1, x_2, p_1, p_2) = \delta(x_1 - x_0) \delta(x_2 - x_0) f(p_1) f(p_2).
\eey  
Then, we compute, $p(M_1) = p(M_2) = w_+ - \frac{1}{2}w_+^2$, and $p(M_3) = (1 - w_+)^2$, where $w_+ = \int_0^{\infty} dp f(p)$ is the fraction of particles with positive momentum.

 \subsection{Stochastic processes }
The analysis of Sec. 3.1 passes with little change to classical stochastic systems. Consider a system characterized by a state space $\Gamma$ with elements $\xi$. Let us denote by $P(\Gamma)$ the space of paths on $\Gamma$, that is, of continuous maps from the time interval $[0, T]$ to $\Gamma$. Here, we are restricting to paths between an initial time $t = 0$, and a final time $t = T$. A stochastic system is described by a probability measure $\mu$ on $P(\Gamma)$, such that the expectation of any function $A$ of $P(\Gamma)$ is given by 
\bey
\langle A\rangle = \int d\mu[\xi(\cdot)] A[\xi(\cdot)]]
\eey

Again, an event $\alpha$ is defined by the first intersection of a path with a surface, and it can be represented by a function $F_{\alpha}$ on $\Gamma$. We can still define a null space $N_{\alpha}$, and a time function $T_{\alpha}$, however, in absence of a deterministic evolution law, these objects are defined on the space of paths $P(\Gamma)$, and not on $\Gamma$. In particular, we define by $N_{\alpha}$ the subset of $P(\Gamma)$ that consists of paths $\xi(\cdot)$ for which the equation $F_{\alpha}(\xi(t)) = 0$ has no solution for any $t \in [0, T]$; we will denote the complement of $N_{\alpha}$ by $\bar{N}_{\alpha}$. We also define the time function $T_{\alpha}$  for any path $\xi(\cdot) \in \bar{N}_{\alpha}$ by setting the value $T_{\alpha}[\xi(\cdot)]$ on a path $\xi(\cdot)$ equal to the smallest value of $t$ such that $F_{\alpha}(\xi(t)) = 0$.

The definition of joint probabilities for the times of events proceeds in a similar way to Sec. 2. For example, the joint probability distribution for $n$ events is
\bey
p(t_1, t_2, \ldots, t_n) = \int d\mu[\xi(\cdot)] \delta(t_1 - T_1(\xi(\cdot)) \delta(t_2 - T_2(\xi(\cdot)) \ldots \delta(t_n - T_n(\xi(\cdot)). 
\eey
The space of paths $P(\Gamma)$ is split into mutually exclusive and exhaustive subsets, each corresponding to an element of $CO(E)$. For two events, we have four elements of $CO(E)$, which correspond to the following subsets.
\bey
M_1 &=&  \{ \gamma \in P(\Gamma)| T_1[\gamma] < T_2[\gamma]\} \cup \left(N_2 \cap \bar{N}_1\right) ,\nonumber \\
M_2 &=&  \{ \gamma \in P(\Gamma)| T_2[\gamma] < T_1[\gamma]\} \cup \left(N_1 \cap \bar{N}_2\right)  ,\nonumber \\
M_3 &=& N_1 \cap N_2, \nonumber \\
M_4 &=& \{ \gamma \in P(\Gamma)| T_1[\gamma] = T_2[\gamma]\} . 
\eey
As an example, consider the case of a Wiener process. We have two particles undergoing Brownian motion, that is, each particle is described by the evolution of a single-time probability density $\rho$ on $\R$, by
\bey
\frac{\partial \rho}{\partial t}  = \frac{D}{2} \frac{\partial^2 \rho}{\partial x^2},
\eey 
where $D$ is the diffusion constant. 
For a particle that starts at $x_0 = -L$, the probability density of crossing the line $x = 0$ is 
\bey
f(t) = \frac{1}{\sqrt{2\pi D t}} \frac{L}{2t} e^{-\frac{L^2}{2Dt}}, \label{fd}
\eey 
with the probability of not crossing $x = 0$ for any $t \in[0, \infty)$ equal to $\frac{1}{2}$.

 We assume that the two particles move independently, and that the associated diffusion constants are different, $D_1$ and $D_2$. (This is possible, for example, if the particle masses are different.)
  The joint probability density that the first crosses $x = 0 $ at time $t_1$ and the second at time $t_2$ is simply $f_1(t_1) f_2(t_2)$, where $f_i$ is the probability density (\ref{fd}), with diffusion constant $D_i$. Then, we evaluate 
\bey
 p(M_1) &=& \frac{1}{2 \pi} \arctan\left(\sqrt{D_1/D_2}\right) + \frac{1}{4}, \nonumber \\
 p(M_2) &=& \frac{1}{2 \pi} \arctan\left(\sqrt{D_2/D_1}\right) + \frac{1}{4}, \nonumber \\
 p(M_3) &=& \frac{1}{4}, 
  \eey
 where we ignored $M_4$, as it is  of measure zero.
 
\section{COoE in quantum systems} 
In this section, we define probabilities for the COoE in quantum systems. 

\subsection{Probability assignment}

For quantum systems, the definition of events in terms of paths crossing a surface does not work, because paths are not physical observables in quantum theory. The only meaningful observables are measurement outcomes. In the most common measurement scheme, namely, von Neumann measurements, the timing of the measurement events is fixed {\em a priori}. Hence, the causal order of events is also fixed.

We need a measurement scheme that treats the time of an event as a random variable, if we are to treat the causal order of events as a random variable quantum mechanically. This is achieved by the QTP approach that was described in the introduction.

Suppose that we have a particle detector located at a fixed region in space. Then, via QTP, we can construct a set of positive $\hat{\Pi}(t)$, such that the probability density of detection at time $t > 0$ is $p(t) = Tr(\hat{\rho}_0\hat{\Pi}(t))$, where $\hat{\rho}_0$ is the initial state if the particle. Together with the positive operator $\hat{\Pi}(N)$ of no detection, the operators $\hat{\Pi}_t$ define a POVM. 

For example, we can construct a POVM for the time of arrival of a particle of mass $m$. We assume that the particle moves at a line and that the particle detector is located at $x = L$. In the momentum basis,
\bey
\langle k| \hat{\Pi}(t)|k'\rangle = \int \frac{dkdk}{2\pi} S(k, k') \sqrt{v_k v_{k'}} e^{i(k-k')L - i(\epsilon_k - \epsilon_{k'})t}, \label{toapovm}
\eey
where $\epsilon_k = \sqrt{m^2+k^2}$ is the particle's energy, $v_k$ is the particle's velocity, and $S(k, k')$ is the {\em localization operator}, that is an operator that determines the irreducible spread of the detection record. The sharpest localization is achieved for $S(k, k') = 1$. The operators $\hat{\Pi}(t)$ are not normalized to unity for $t \in[0, \infty)$. However, if we restrict to quantum states with strictly positive momentum content, the contribution to the total probability from negative values of $t$ is negligible, and we can consider the normalization of $\hat{\Pi}(t)$ in the full real line. In this case,
\bey
\int_{-\infty}^{\infty} dt \hat{\Pi}(t) = \hat{I}.
\eey

 For $n$ independent detectors, each detecting a different particle, we can identify a POVM $\hat{\Pi}(t_1, t_2, \ldots, t_n) = \hat{\Pi}_1(t_1) \otimes \hat{\Pi}_2(t_2) \otimes \ldots \otimes \hat{\Pi}_n(t_n)$, where $\hat{\Pi}_i(t_i)$ corresponds to the POVM for the $i$-th detector. Thus, we can define probability densities $p(t_1, t_2, \ldots, t_n)$ for the $n$ measurement events, and we can follow the same procedure  as in Sec. 3, in order to obtain probabilities for different causal orders of $n $ measurement events. 
 
 For example, for two events, $1$ and $2$, we have the three COoEs $M_1 = \{1 \po 2\}$, $M_2 = \{2 \po 1\}$, and $M_3 = \{1 ||2\}$. The positive operators 
 \bey
 \hat{E}(M_1) &=& \int_0^{\infty} dt_2 \int_0^{t_2} dt_1 \hat{\Pi}_1(t_1) \otimes \hat{\Pi}_2(t_2) + [\hat{I} - \hat{\Pi}_1(N)] \otimes \hat{\Pi}_2(N), \nonumber  \\ 
 \hat{E}(M_2) &=& \int_0^{\infty} dt_1 \int_0^{t_1} dt_2 \hat{\Pi}_1(t_1) \otimes \hat{\Pi}_2(t_2) + \hat{\Pi}_1(N) [\hat{I} - \hat{\Pi}_2(N)], \nonumber \\ 
  \hat{E}(M_3) &=& \hat{\Pi}_1(N) \otimes \hat{\Pi}_2(N),
 \eey
 define a POVM for the causal orders. 
 
\begin{figure}
 \includegraphics[width=5cm]{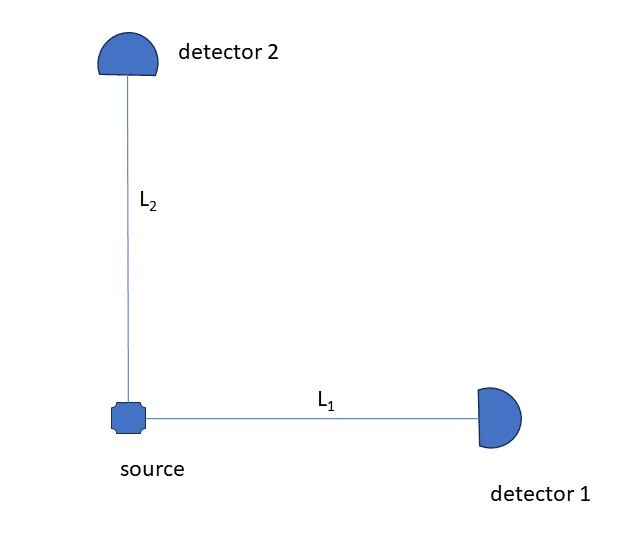} 
    \caption{ A set-up by which to measure the causal order of two events that correspond to detections at detectors 1 and 2. }
    \label{ills}
\end{figure}

Assume that the two events correspond to the detection of identical particles with the two detectors located at $x_1 = L_1$ and $x_2 = L_2$ from the source---see Fig. \ref{ills}. We use the POVM (\ref{toapovm}) for both $\hat{\Pi}_1(t)$ and $\hat{\Pi}_2(t)$.  Taking $- \infty$ for the lower bound in the time integral, we find
\bey
\hat{E}(M_1) &=& \frac{1}{2} \hat{I} + \hat{B}\\
\hat{E}(M_2) &=& \frac{1}{2} \hat{I} -  \hat{B}\\
\hat{E}(M_3) &=& 0
\eey
 where the operator $\hat{B}$ is defined by matrix elements
 \bey
 \langle k_1, k_2|\hat{B}|k_1', k_2'\rangle &=& i S_1(k_1, k_1') S_2(k_2, k_2') \sqrt{v_{k_1}v_{k_2}v_{k_1'}v_{k_2'}} \nonumber \\
 &\times& e^{i(k_1 - k_1')L_1 - i (k_2 - k_2')L_2} \delta(\epsilon_{k_1} + \epsilon_{k_2} - \epsilon_{k_1'}-\epsilon_{k_2'}) \; \text{PV}\frac{1}{\epsilon_{k_2} - \epsilon_{k_2'}}.  
 \eey
 Here, PV stands for Cauchy principal value.
 
 Consider a general initial state of the form $|\psi\rangle = \sum_i c_i |\psi_{1i}\rangle \otimes |\psi_{2i}\rangle$. The probability densities associated to the three orders are
 \bey
 p(M_1) = \frac{1}{2} + w, \ \ p(M_2) = \frac{1}{2} - w, \ \ p(M_3) = 0,
 \eey
 where the asymmetry
 \bey
 w = \sum_{ij} c_i c^*_j \int_{-\infty}^{\infty} ds  \ [\alpha^{(1)}_{ij}(s) \dot{\alpha}^{(2)}_{ij}(s) - \dot{\alpha}^{(1)}_{ij}(s) \alpha^{(2)}_{ij}(s)], \label{ww}
 \eey
 is expressed in terms of the quantities 
 \bey
 \alpha_{ij}^{(a)}(s) = \int \frac{dkdk'}{2\pi} \psi_{ai}(k) \psi_{aj}^*(k') S_a(k, k') \sqrt{v_k v_{k'}} e^{i(k - k')L_a - i (\epsilon_k - \epsilon_{k'})s}\ \text{PV} \frac{1}{\epsilon_{k} - \epsilon_{k'}},
 \eey 
where $a  = 1, 2$.  

\subsection{Examples}

\begin{figure}
 \includegraphics[width=0.34\textwidth]{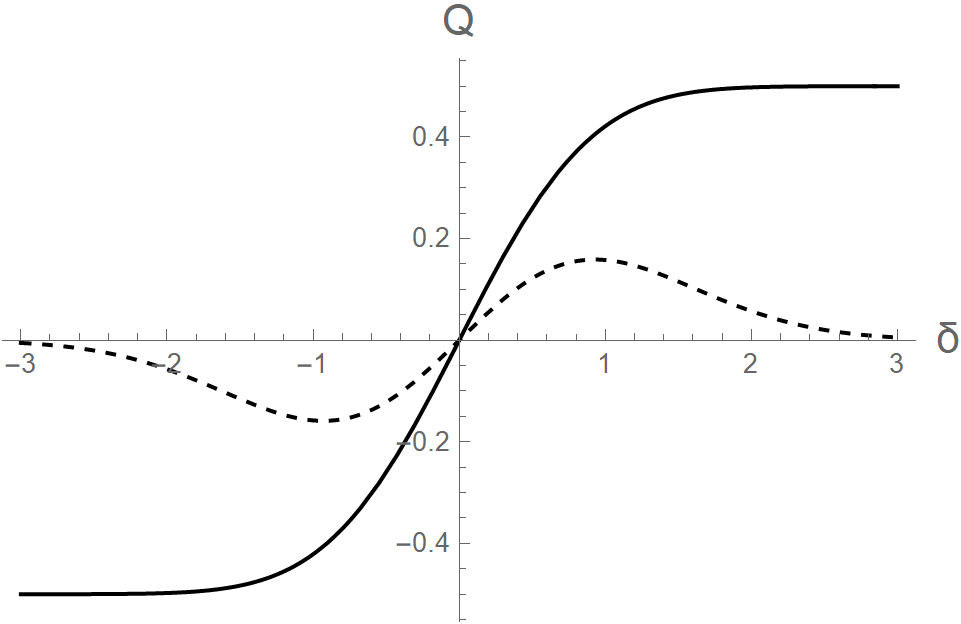} 
    \caption{ The function $Q_1$ of Eq. (\ref{www}) (solid) and the function $Q_2$ of Eq. (\ref{Q2}) (dashed) as a function of $\delta$. }
    \label{delt}
\end{figure}

We analyze the case of massless particles, $m = 0$, and ideal detector, $S(k,k') = 1$. For two particles prepared in the same state $\psi_0(k)$, that is centered around $x = 0$. However, the distances traveled by the two particles may be different, $L_1 \neq L_2$. We take for $\psi_0$ a Gaussian centered around $k_0$, 
\bey
\psi_0(k) = (2 \pi \sigma)^{-1/4} \exp\left[(k-k_0)^2/(4\sigma^2)\right].\label{psi0}
\eey
 Then, we find that $w$ in Eq. (\ref{ww}) equals  $Q_1[\sigma(L_1 - L_2)]$, where 
\bey
Q_1(\delta) = \int_{-\infty}^{\infty} dx e^{-2 (x - \delta)^2} \mbox{erf}\left(\sqrt{2}x\right). \label{www}
\eey
The dependence of the function $Q$ on $\delta$ is plotted in Fig. \ref{delt}. As expected $w$ vanishes for $L_1 = L_2 $ and tends to $\pm \frac{1}{2}$ for large differences between $L_1$ and $L_2$, in which case the ordering of the events is almost certain.

The probabilities in the example above could have been derived from a classical theory. To see quantum behavior, we consider an superposition state for the first particle 
\bey
\psi(k_1, k_2) = \frac{1}{\sqrt{2(1 + \nu)}} \psi_0(k_1) \left[1 + e^{ik_1\ell}\right]\psi_0(k_2),  
\eey 
 Here $\ell$ is a path difference for the first particle in one component of the superposition and $\nu = \int dk_1 dk_2   |\psi_0(k)|^2 \cos(k\ell)$. For the Gaussian (\ref{psi0}), $\nu = e^{-\sigma^2\ell^2/2}\cos(k_0\ell)$. For this initial state
\bey
w = \frac{Q_1(\delta) + 2Q_2(\delta) \cos\left(\frac{k_0}{\sigma}\delta\right)}{2\left[1 +e^{-\delta^2/2}\cos\left(\frac{k_0}{\sigma}\delta\right)\right]}, \label{eeee}
\eey
where $\delta = \sigma \ell$, and the function
\bey
Q_2(\delta) = \int_{-\infty}^{\infty} dx e^{-x^2 - (x - \delta)^2} \mbox{erf}\left(\sqrt{2}x\right), \label{Q2}
\eey 
is plotted in Fig. \ref{delt}. 

\begin{figure}
 \includegraphics[width=0.8\textwidth]{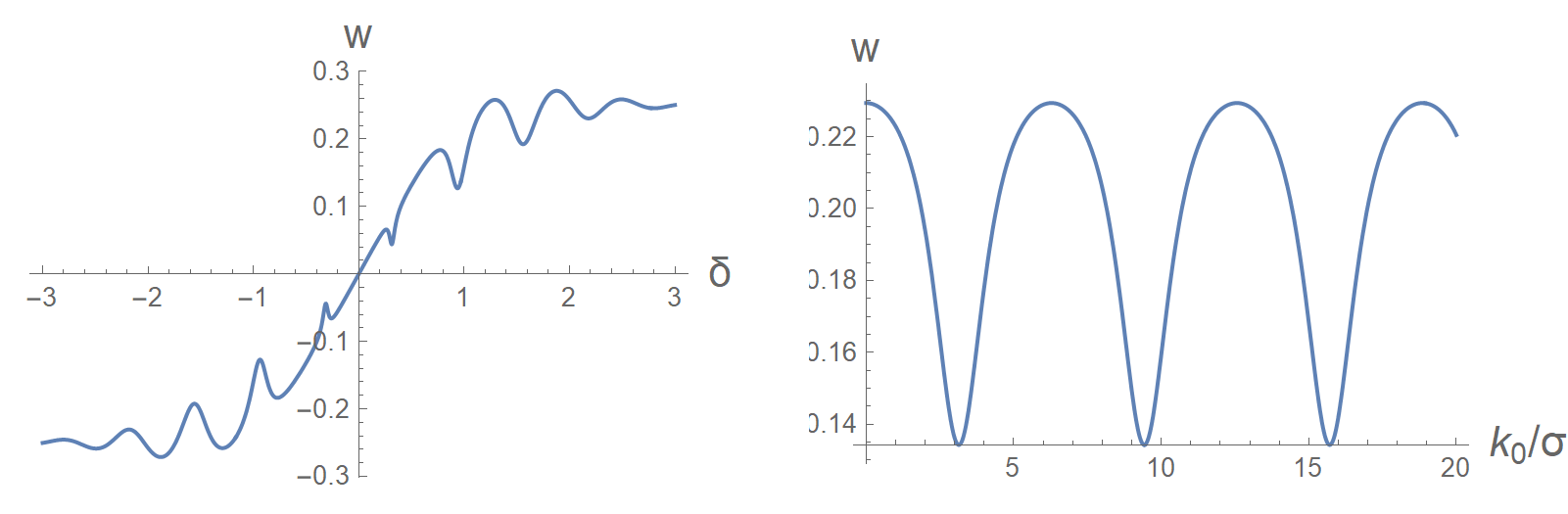} 
    \caption{ The asymmetry  $w$ of Eq. (\ref{eeee}) as a function of $\delta$ for constant 
  $k_0/\sigma = 10$  (left) and as a function of $k_0/\sigma$ for constant $\delta = 1$ (right).  }
    \label{wplot}
\end{figure}

In Fig. (\ref{wplot}), we plot $w$ as a function of $\delta$ and of $k_0/\sigma$. The quantum nature of the system is manifested in the oscillatory behavior of the probabilities.
 \section{Probabilities for the COoE via a detection model}
 
In the examples of Sec. 4, the measurements of the causal ordering of events are coarse-grained. This means that  the measuring apparatuses record the time of  detection events, and the probabilities for the causal order of events are obtained by integrating over probabilities with respect to detection times. In this sense, the construction is formally similar to the one of classical physics, even if there are no paths at the fundamental level.

However, in quantum theory, it may be possible to define probabilities for causal ordering of events as fine-grained observables, even if we cannot distinguish between the times of the individual events. In this section, we will present a simple model that provides such probabilities for the case of two potential events, and which can be straightforwardly generalized for $n$ events.

\subsection{The model}
The key idea is to direct a pair of particles towards a detector that can record only one of them, but not both. As an example, we consider a three-level system (3LS), with states $|0\rangle, |1\rangle$, and $|2\rangle$. Suppose that particle 1 can excite only the transition $0 \rightarrow 1$, and particle 2 only the transition $0 \rightarrow 2$. If after the interaction of the particles with the three-level system, we find the system in state $|1\rangle$, we can surmise that particle 1 was detected first and particle 2 was not detected, and vice versa.  

To conform with our definition of an event with a measurement record, we must place an identical 3LS after the first, in which the particle not absorbed by the first can be detected. However, this is superfluous for identifying the COoE in this system, so we will consider the case of a single 3LS.

This set-up is straightforwardly generalized for determining the probabilities for $n$ events. We require $n$ particles that can be sharply distinguished by  their energies and $n - 1$ systems with $n + 1$ energy levels, so that the detection of each particle can be associated to a single transition.

To implement our model, we assume that the incoming particles are described by a free scalar field $\hat{\phi}(x)$ with mass $m$. The two particles are distinguished by their initial energies; we can assume that they are prepared from different sources. The particles interact with one 3LS, which we take to be located at ${\bf x} = 0$.

  The total Hamiltonian is a sum of three terms $\hat{H}_{\phi} + \hat{H}_{3LS} + \hat{H}_{int}$, where $\hat{H}_{\phi}$ is the field Hamiltonian, expressed in terms of field creation and annihilation operators,
\bey
\hat{H}_{\phi} = \int d{\bf k} \epsilon_{\bf k} \hat{a}^{\dagger}_{\bf k} \hat{a}_{bf k},
\eey
where $d {\bf k} $ stands for $d^3k/(2\pi)^3$; 
\bey
\hat{H}_{3LS} = \Omega_1 |1\rangle \langle 1| + \Omega_2 |2\rangle \langle 2|
\eey
is the 3LS Hamiltonian, and the interaction Hamiltonian
\bey
\hat{H}_{int} = \sum_{a=1}^2\lambda_a \int \frac{dk}{\sqrt{2\omega_{\bf k}}} (\hat{a}_{\bf k} \hat{u}_{a+} + \hat{a}^{\dagger}_{\bf k} \hat{u}_{a-}),
\eey
describes a dipole coupling between the field and the 3LS. Here, 
 $\lambda_a$ is the coupling constants associated to the transition $0 \rightarrow a$, $\hat{u}_{a+} = |a\rangle \langle 0|$, and $\hat{u}_{a-} = |0\rangle \langle a|$; $a = 1, 2$.  This Hamiltonian is a variation of Lee's Hamiltonian that is commonly employed in the study of spontaneous decay \cite{Ana23}.

\subsection{Time evolution} 
 
To derive the time evolution law for this model, we work in the interaction picture. Then,  
the quantum state satisfies the equation
\bey
i \frac{\partial}{\partial t} |\psi(t) \rangle =  \sum_{a=1}^2\lambda_a \int \frac{d{\bf k}}{\sqrt{2\omega_{\bf k}}} (\hat{a}_{\bf k} \hat{u}_{a+} e^{-i (\epsilon_{\bf k} - \Omega_a)t} + \hat{a}^{\dagger}_{\bf k} \hat{u}_{a-}e^{i (\epsilon_{\bf k} - \Omega_a)t})|\psi(t)\rangle. \label{evolution}
\eey
 We assume an initial two-particle state for the field and the ground state for the 3LS. The Hamiltonian employed here causes transitions only to one-particles states with an excited state for the 3LS. Hence, the state is of the form
 \bey
 |\psi(t) \rangle = \int d{\bf k} d{\bf k}' c({\bf k}, {\bf k}'; t) |{\bf k}, {\bf k}',0\rangle + \sum_a \int d{\bf k} d_a({\bf k}; t) |k, a\rangle,
 \eey
Substituting into Eq. (\ref{evolution}), we obtain
\bey
i \dot{c}({\bf k}, {\bf k}'; t) &=& \sum_a  \lambda_a \left[\frac{d_a({\bf k}; t) }{\sqrt{2\epsilon_{{\bf k}'}}} e^{i(\epsilon_{{\bf k}} - \Omega_a)t} + \frac{d_a({\bf k}'; t) }{\sqrt{2\epsilon_{\bf k}}} e^{i(\epsilon_{{\bf k}'} - \Omega_a)t} \right] \label{ev1}\\
 i\dot{d}_a({\bf k}; t) &=& 2 \lambda_a \int \frac{d{\bf k}'}{\sqrt{2 \epsilon_{{\bf k}'}}} c({\bf k}, {\bf k}'; t)e^{-i(\epsilon_{{\bf k}'} - \Omega_a)t} \label{ev2}
\eey
These equations are to be solved subject to the initial conditions $d_a({\bf k}; 0) = 0$ and $c({\bf k}, {\bf k}'; 0) = c_0({\bf k}, {\bf k}')$, where $c_0({\bf k}, {\bf k}')$ is the initial state of the two particles. We integrate both sides of Eq. (\ref{ev1}) and substitute $c({\bf k}, {\bf k}'; t)$ to Eq. (\ref{ev2}). We obtain
\bey
\dot{d}_a({\bf k}; t) &=&  2 \lambda_a \int \frac{d{\bf k}'}{\sqrt{2\epsilon_{{\bf k}'}}} c_0({\bf k}, {\bf k}')e^{-i(\epsilon_{{\bf k}'} - \Omega_a)t}\nonumber \\
&-& 2 \lambda_a \sum_b \lambda_b \int \frac{d{\bf k}'}{2\epsilon_{{\bf k}'}}e^{-i(\epsilon_{{\bf k}'} - \Omega_a)t} \int_0^t ds d_a({\bf k};s) e^{i(\epsilon_{{\bf k}} - \Omega_b)s} \nonumber \\
&-& 2 \frac{\lambda_a}{\sqrt{\epsilon}_{\bf k}} \sum_b \lambda_b \int \frac{d{\bf k}'}{2\sqrt{\epsilon_{{\bf k}'}}}e^{-i(\epsilon_{{\bf k}'} - \Omega_a)t} \int_0^t ds d_a({\bf k}';s) e^{i(\epsilon_{{\bf k}'} - \Omega_b)s}. \label{intdif}
\eey
Eq. (\ref{intdif}) is exact.  The term in the second line is proportional to the vacuum Wightman function $W(t) =  \int \frac{d{\bf k}'}{2\epsilon_{{\bf k}'}}e^{-i\epsilon_{{\bf k}'}t}$, which drops at least with $e^{-m t}$ for $m \neq 0$ and as $t^{-2}$ for $m = 0$. Assuming that the particle starts sufficiently far from the detector, $d_a({\bf k}; t)$ becomes appreciable different from zero at times such that the term proportional to $W(t)$ is strongly suppressed. 

The third-line term in Eq. (\ref{intdif}) is of a structure that commonly appears in elementary treatments of spontaneous decay \cite{QOp, Ana23}. It can be calculated by invoking a version of the Wigner-Weisskopf approximation. For $\Omega_a t >> 1$, this expression is strongly dominated by the term with $b = a$. By carrying out the integration over ${\bf k}'$, we obtain
\bey
- \frac{\lambda_a^2}{2 \pi^2\sqrt{\epsilon_{\bf k}}} \int_m^{\infty} d \epsilon  \frac{(\epsilon^2 - m^2)^{3/2} }{\sqrt{\epsilon}} \int_0^t ds e^{- i (\epsilon - \Omega_a)(t - s)} d_a({\bf k}, s). \label{term3}
\eey
The time integral is negligible except for values of $\epsilon$ around $\Omega_a$. Hence, we are justified in substituting $(\epsilon^2 - m^2)^{3/2}/\sqrt{\epsilon}$ with $(\Omega_a^2 - m^2)^{3/2}/\sqrt{\Omega_a}$, and then, to extend integration over $\epsilon $ to $(-\infty, \infty)$. Then, the term (\ref{term3}) simplifies to $-\frac{1}{2} \eta_a \epsilon_{\bf k}^{-1/2} d_a({\bf k}, t)$, where
\bey
\eta_a = - \frac{\lambda_a^2 (\Omega_a^2 - m^2)^{3/2}}{\pi\sqrt{\Omega_a}}. 
\eey
 Eq. (\ref{intdif}) becomes
 \bey
 \dot{d}_a({\bf k}; t) + \frac{1}{2} \eta_a \epsilon_{\bf k}^{-1/2} d_a({\bf k}; t) =  2 \lambda_a \int \frac{d{\bf k}'}{\sqrt{2\epsilon_{{\bf k}'}}} c_0({\bf k}, {\bf k}')e^{-i(\epsilon_{{\bf k}'} - \Omega_a)t}.
 \eey
 This is a linear inhomogenous equation of first order. The Green function for the corresponding homogeneous equation is simply $\theta (t - t') e^{-\frac{1}{2}\eta_a \epsilon_{\bf k}^{-1/2}(t - t')}$. Hence, we obtain
 \bey
 d_a({\bf k}; t) = 2 \lambda_a  \int \frac{d{\bf k}'}{\sqrt{2\epsilon_{{\bf k}'}}} c_0({\bf k}, {\bf k}')\int_0^t ds e^{-\frac{1}{2}\eta_a \epsilon_{\bf k}^{-1/2}(t - s)}e^{-i(\epsilon_{{\bf k}'} - \Omega_a)s} \nonumber =
 2 \lambda_a \int d{\bf k}'c_0({\bf k}, {\bf k}') h_a(\epsilon_{\bf k}, \epsilon_{{\bf k}'};t), 
 \eey 
where 
\bey
h_a(\epsilon, \epsilon'; t) = \frac{e^{-\frac{1}{2}\eta_a \epsilon^{-1/2} t} - e^{-i(\epsilon' - \Omega_a)t}}{\sqrt{2\epsilon'}\left[\frac{1}{2}\eta_a \epsilon^{-1/2} - i (\epsilon' - \Omega_a)\right]}.
\eey
The detection probability is non-negligible only if ${\bf k}$ is along the axis that connects the source to the detector. Hence, the problem is effectively one-dimensional. Therefore, we can substitute the initial state with $c_0(k, k')$, where $k, k' > 0$, and write $
d_a(k; t) = 2 \lambda_a  \int \frac{dk'}{2\pi} c_0(k, k') h_a(\epsilon_k, \epsilon_{k'};t)$.

\subsection{An example} 
\begin{figure}
 \includegraphics[width=0.34\textwidth]{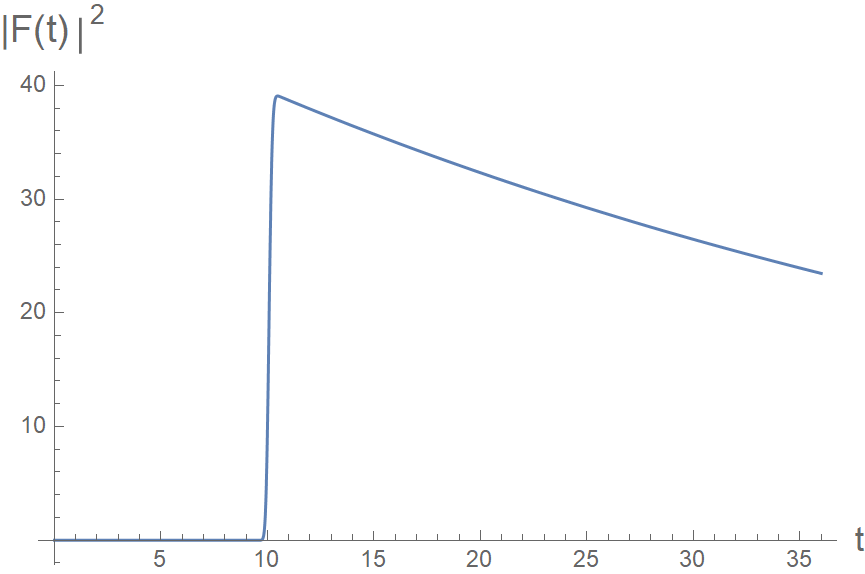} \hspace{1cm}
    \caption{Typical plot of the functions $|F_{ia}(t)|^2$ as a function of time $t$, for a Gaussian function $\chi(k)$. The function jumps to a finite value when the particle arrives at the 3LS, and  then it decays with a rate of $\Gamma_{ia}$. }
    \label{fia}
    \label{fig:foobar}
\end{figure}

Consider an initial state 
\bey
c_0(k, k') = \frac{1}{\sqrt{2}} \left[\psi_1(k) \psi_2(k') + \psi_1(k') \psi_2(k)\right], 
\eey
where $\psi_i$, for $i = 1, 2$, is centered around momentum $k_i$, or, equivalently, on energy $\epsilon_i = \sqrt{k_i^2 +m^2}$. We assume that there is no overlap between $\psi_1$ and $\psi_2$. Then, we can approximate
\bey
d_a(k; t) = \lambda_a \left[\psi_1(k) F_{2a}(t) + \psi_2(k) F_{1a}(t) \right],
\eey
where
\bey
F_{ia}(t) = \int \frac{dk}{2\pi} \psi_i(k) \frac{ e^{-\Gamma_{ia}t} - e^{-i(\epsilon_k - \Omega_a)t}}{\sqrt{2\epsilon_k}\left[\Gamma_{ia} - i (\epsilon_k - \Omega_a)\right]},\\
\eey
where $\Gamma_{1a} = \frac{1}{2}\eta_a \epsilon_2^{-1/2}$ and  $\Gamma_{2a} = \frac{1}{2}\eta_a \epsilon_1^{-1/2}$. 
Then, the probability $p_a(t)$ that the 3LS is found in an excited state is given by
\bey
p_a(t) = \int dk |d_a(k; t)|^2 = \lambda_a^2 \left( |F_{1a}(t)|^2 + |F_{2a}(t)|^2\right).
\eey
 Let the states $\psi_i(k)$ be well localized around $x = - L$, so that they can be written as $\chi(k - k_i) e^{ikL}$, where $\chi$ is a positive function peaked around $k = 0$, for example, a Gaussian. Then, the typical behavior of $|F_{ia}(t)|^2$ is given in Fig. \ref{fia}. The function is negligible prior to the arrival time $t_a = mL/k_i$ of the particle to the locus of 3LS. Then, it jumps to a finite value, which then decays with a rate given by $\Gamma_{ia}$.

 The peak value of $|F_{ia}(t)|^2$ is approximately proportional to the Breit-Wigner term $[(\Gamma_{ia}^2 + (\epsilon_i - \Omega_a)^2]^{-1}$. Supposing that we choose $\epsilon_a \simeq \Omega_a$, and that $\Gamma_{ia} << |\Omega_1 - \Omega_2|$, for all $i, a = 1, 2$, then, the terms $|F_{11}|^2$ and $|F_{22}|^2$ dominate in the probability assignment, and 
\bey
p_a(t) = \lambda_a^2 |F_{aa}(t)|^2.
\eey
The behavior of the probabilities is characteristic of resonant fluorescence. The 3LS absorbs one of the two particles, and after a time of order $\Gamma_{aa}^{-1}$ it re-emits the particle, albeit in a different direction. Hence, the energy of the fluorescent particle determines whether the ordering $M_1$ or $M_2$ was realized.


\section{Conclusions}
We provided a general definition of events in quantum theory, and showed how to construct probabilities associated to the causal ordering such events. Our notion of events is very different from that of Refs. \cite{OCB12, CDPV13}, and it is naturally related to the relativistic notion of events. Our analysis clarifies that the existence of an indefinite quantum causal order of events has no relation to quantum gravity, as this causal order is a dynamical consequence of the quantum nature of the {\em matter} degrees of freedom. The COoE should not be conflated with the causal structure of spacetime, which we take to be fixed and unchanged in absence of gravity. Further work is needed, in order to explore how the quantum probabilities for causal order defined here differ from the corresponding classical one, for example, if they violate Bell' type inequalities.

The model systems considered in this paper are experimentally accessible. The set-ups considered in Sec. 5 are essentially quantum races, that is the causal order of events coincides with the order that a number of distinguishable particles arrive in a specific finish line. The set-up of Sec. 6, when applied to photons, involves a variation of resonant fluorescence with specially engineered  multi-level atoms that play the role of detectors for the causal ordering that is being realized.

\end{document}